\documentclass[twocolumn,prl,showpacs]{revtex4}

\usepackage{graphicx}
\usepackage{dcolumn}
\usepackage{float}
\usepackage{amsmath}

\newcommand{\comment}[1]{}

\begin{document}


\title{High energy scales in electronic self-energy imaged by optics}
\author{J. Hwang$^1$}
\email{hwangjs@mcmaster.ca}

\author{E.J. Nicol$^2$}
\email{nicol@physics.uoguelph.ca}

\author{T. Timusk$^{1,3}$}

\author{A. Knigavko$^{4}$}

\author{J.P. Carbotte$^{1,3}$}

\affiliation{$^1$Department of Physics and Astronomy, McMaster
University, Hamilton, Ontario N1G 2W1 Canada}
\affiliation{$^2$Department of Physics, University of Guelph,
Guelph, Ontario N1G 2W1 Canada} \affiliation{$^{3}$The Canadian
Institute of Advanced Research, Toronto, Ontario M5G 1Z8, Canada}
\affiliation{$^{4}$ Department of Physics, Brock University, St.
Cathrine, Ontario L25 3A1, Canada}

\date{\today}

\begin{abstract}
We use a new technique to directly extract an estimate of the
quasiparticle self-energy from the optical conductivity which can
be easily related to both theory and angle-resolved photoemission
spectroscopy (ARPES) experiments. In the high $T_c$ cuprate
Bi-2212 we find evidence for a new high energy scale at 900 meV in
addition to the two previously well known ones at roughly 50 and
400 meV. The intermediate scale at 400 meV has been recently seen
in ARPES as a large kink which optics finds to be weaker and
shifted. In YBCO, the three energy scales are shifted to lower
energy relative to Bi-2212 and we observe the emergence of a
possible fourth high energy feature at 600 meV.
\end{abstract}
\pacs{74.25.Gz,74.72.-h,74.20.Mn}

\maketitle

One of the most beautiful aspects of studying condensed
matter systems is that a large number of very different
types of experiments are available to probe the many-body
state, and in the quest for microscopic understanding
all of the various probes must concur about the features
observed. This provides a stringent test of theories
and further allows for the perfection of experimental
technique and analysis. Consequently, if there is an
opportunity to bring two different techniques closer
together for the purpose of comparison, both theory and
experiment will benefit and there will be a great enhancement
of the ability to identify potentially fundamental and robust
signatures of the underlying physics. Two powerful
techniques which  have been brought to bear on the problem of
high $T_c$ superconductivity are angle-resolved photoemission
spectroscopy (ARPES)\cite{campuzano,damascelli}
and far-infrared optical spectroscopy\cite{basovrmp}.
These techniques have blossomed with the demands of this
area of study resulting in unprecedented resolution and
new methods for extraction and analysis of information,
such as through inversion techniques.\cite{zhou,verga,marsiglio,
dordevic,schachinger,carbnat}

Applying inversion techniques to the optical conductivity has
resulted in the identification of a coupling of the charge
carriers to a resonance mode at about 50 meV and a spin
fluctuation background extending beyond 400
meV.\cite{schachinger,carbnat,timusk,hwang06} ARPES, on the other
hand, has mainly confirmed the low energy scale as, until recently,
they were limited to presenting data for the quasiparticle
dispersion curve up to about 200-300 meV in
energy.\cite{kaminski,kaminski2,sato,
kordyukmag,kordyuk2,kim,johnson,lanzara,cuk} However, new advances
have extended this range to greater than 1 eV and several
works\cite{graf,xie,vallas06,graf06,pan06} are reporting features
at high energy at about 400 meV and 800 meV, which are speculated
to arise from exotic physics. Confirming the existence of these
new energy scales and investigating a wide range of materials  is
essential to the development of these or other new ideas.

In this letter, we present complementary results obtained from
optical data and demonstrate, at the same time, a new technique
which allows for the direct, although approximate, extraction of
the {\it quasiparticle} self-energy from the optical conductivity,
which can be compared most easily with the ARPES via a
dispersion-type curve. This is important because optics is a bulk
probe as opposed to ARPES which is only sensitive to the surface.
In addition, optical spectroscopy measures the self-energy due to
correlations directly while ARPES finds, instead, the renormalized
energy and the self-energy follows only once the bare dispersion
is known. Using our new technique to analyze the data on Bi-2212
samples of varying doping, we can show that optics also sees
features at high energy and supports some of the observations seen
in ARPES, but not all. Moreover, as optical experiments are more
flexible in the type of materials which can be studied, our work
goes beyond Bi-2212 to examine the high energy scales in YBCO.

We begin by discussing the theoretical basis for the new analysis
of optics which allows for the extraction of the quasiparticle
self-energy. Within the experimental community, it has become
common to analyze the complex optical conductivity
$\sigma(\omega)$ for the correlated charge carriers in terms of
the optical self-energy,
$\Sigma^{op}=\Sigma^{op}_1(\omega)+i\Sigma^{op}_2(\omega)$, using an
extended Drude model of the form:\cite{timusk,basovrmp}
\begin{eqnarray}
\sigma(\omega)&=&i\frac{\omega_p^2}{4\pi}\frac{1}{\omega[1+\lambda^{op}(\omega)]
+i/\tau^{op}(\omega)}\label{eq:sig1}\\
&=&i\frac{\omega_p^2}{4\pi}\frac{1}{\omega-2\Sigma^{op}(\omega)}
\label{eq:sig2}
\end{eqnarray}
where $\omega_p$ is the plasma frequency. $\Sigma^{op}$ can be
written in terms of the optical scattering rate
$1/\tau^{op}(\omega)$ and the optical mass renormalization
$1+\lambda^{op}(\omega)$ by
\begin{eqnarray}
\frac{1}{\tau^{op}(\omega)} &=&\frac{\omega_p^2}{4\pi}
{\rm Re}\biggl(\frac{1}{\sigma(\omega)}\Biggr)=
-2\Sigma_2^{op}\label{eq:tau}\\
\omega\lambda^{op}(\omega) &=&-\frac{\omega_p^2}{4\pi}
{\rm Im}\Biggl(\frac{1}{\sigma(\omega)}\Biggr)-\omega =
-2\Sigma_1^{op}
\label{eq:lam}
\end{eqnarray}
and it is routine to present quantities such as $1/\tau^{op}$ or
$\Sigma^{op}$. Despite the suggestive structure of this form,
$\Sigma^{op}$ is not the quasiparticle self-energy $\Sigma^{qp}$
and $1/\tau^{op}$ is not the quasiparticle scattering rate.
However, these quantities do encode such information and within a
set of reasonable approximations, a surprisingly simple procedure
can be applied to extract $\Sigma^{qp}$ from $\Sigma^{op}$.

If we can assume, for the sake of argument, that the
renormalization and scattering of charge carriers is due to some
sort of frequency-dependent interaction with a bosonic spectrum
which we will call $\alpha^2F(\omega)$ in analogy to the
conventional electron-phonon spectral function, then we can show
that\cite{antonclean,antonimp,cappelluti,boza}
\begin{eqnarray}
\frac{1}{\tau^{op}} &\simeq&\frac{2\pi}{\omega}\int_0^\infty\, d\Omega\alpha^2F
(\Omega)\int_0^{\omega-\Omega}\, d\omega'\tilde N(\omega')
\label{eq:tauapprox}\\
\omega\lambda^{op}(\omega) &\simeq&\frac{2}{\omega}\int_0^\infty\,
d\Omega\alpha^2F(\Omega)\nonumber\\
&&\times\int_0^\infty\, d\omega'\tilde N(\omega')
\ln\Biggl(\frac{[\omega'+\Omega]^2}
{[\omega'+\Omega]^2-\omega^2}\Biggr)
\label{eq:lamapprox}
\end{eqnarray}
where the electronic density of states $\tilde N(\omega)$ can be
energy-dependent and we have ignored the difference between the
transport and quasiparticle spectral densities, $\alpha_{tr}^2F$
and $\alpha^2F$, respectively. The approximations behind these
formulas and their range of validity may be found in the work of
Allen\cite{allen}, Shulga et al.\cite{shulga}, and Sharapov and
Carbotte\cite{sharapov}. For convenience, these equations are
stated here for $T=0$ and for no impurity scattering. In the
cuprates, we can assume the clean limit. Finite temperatures
effects are easily incorporated when needed. If we define a new
quantity, following the work of Carbotte et al.\cite{carbotte},
which we will call the optically-derived quasiparticle
self-energy,
\begin{equation}
\Sigma^{op-qp}(\omega)
\equiv\frac{d}{d\omega}[\omega\Sigma^{op}(\omega)]
\label{eq:sigopqp}
\end{equation}
then from Eqs.~(\ref{eq:tauapprox}) and (\ref{eq:lamapprox}), the
derivative in Eq.~(\ref{eq:sigopqp}) gives exactly $\Sigma^{qp}$.
%
%
\begin{figure}[t]
  \vspace*{-0.8 cm}%
  \centerline{\includegraphics[width=3.30 in]{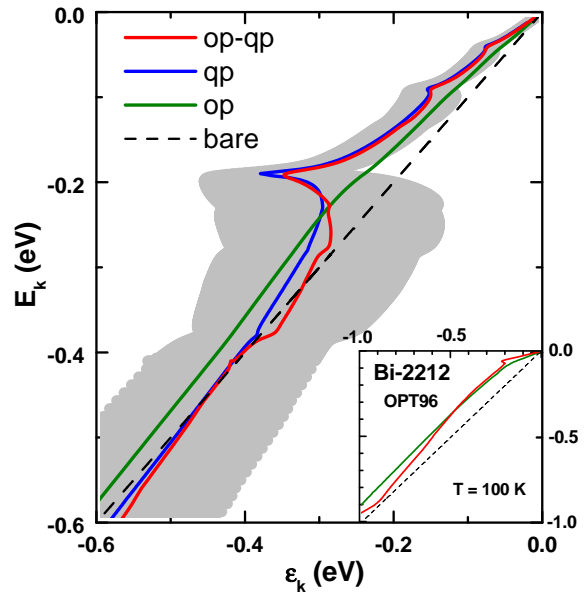}}%
  \vspace*{-1.8 cm}%
\caption{(color online) Theoretical illustration based on the
model described in the text which demonstrates that the
quasiparticle dispersion curve can be extracted from optics
(op-qp) via our new procedure. It agrees well in magnitude and
detailed structure with the exact theoretical curve (qp). If the
optical self-energy is used, the resulting dispersion curve (op)
is structureless and does not agree with the exact result. The
inset illustrates for a real material Bi-2212 the difference
between the dispersion curve derived from the optical self-energy
and that from the new procedure which gives rise to detailed
structure and specific energy scales.}
  \label{Fig1}
\end{figure}
%
%
\begin{figure}[t]
  \vspace*{-2.3 cm}%
  \centerline{\includegraphics[width=3.30 in]{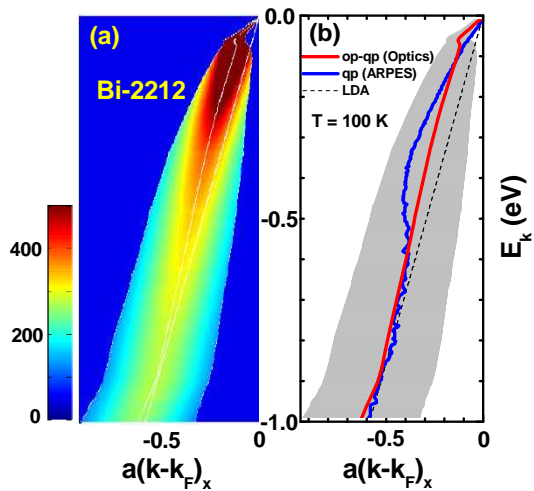}}%
  \vspace*{-2.3 cm}%
\caption{(color online) (a) Dispersion curve derived from optics
for optimally-doped Bi-2212, for $T = 100$ K, in a color map
showing the relative height and width of $A(k,w)$, as is done in
ARPES. (b) Dispersion curve from (a) repeated (op-qp) and compared
with the result reported by ARPES and with the theoretical LDA
curve from Ref.~\cite{bansil}. The grey shading gives the width of
the optically-derived quasiparticle curve based on the imaginary
part of the self-energy.}
  \label{Fig2}
\end{figure}

To illustrate the effectiveness of Eq.~(\ref{eq:sigopqp}) to
determine the exact quasiparticle self-energy, we show a model
calculation in Fig.~\ref{Fig1}, where an $\alpha^2F(\omega)$
spectrum consisting of three delta functions has been used to
calculate both the exact $\Sigma^{qp}$ from many-body theory and
the optical conductivity in the usual Kubo
approximation.\cite{antonclean,antonimp} The optical conductivity
is then processed via Eqs.~(\ref{eq:tau})-(\ref{eq:lam}) and
(\ref{eq:sigopqp}) to obtain $\Sigma^{op-qp}$.  In the figure, we
have chosen to plot the real part of $\Sigma$ in terms of a
renormalized dispersion curve where
$E_k=\epsilon_k-\Sigma_1(E_k)$~\cite{schachinger03} as we wish
eventually to make comparison with ARPES in the same format. The
dispersion curve for $\Sigma^{qp}$ shows very clearly three
structures due to the three $\delta$-function peaks in the
$\alpha^2F$ at 0.04, 0.09, and 0.19 eV\cite{antonimp}. However,
the curve formed from $\Sigma^{op}$ is fairly featureless and does
not capture any of the underlying structure of the spectral
function. In addition, this calculation has included a finite band
density of states\cite{antonclean,antonimp,cappelluti} which
causes the renormalized dispersion curve to cross the bare
dispersion curve (a fact that is not well-known, but is a
signature of finite bands) and once again the curve from
$\Sigma^{op}$ has not crossed the bare dispersion at the same
energy and indeed the crossing is higher in energy by a factor of
roughly two. On the other hand, the optically-derived
quasiparticle self-energy $\Sigma^{op-qp}$ gives a dispersion
curve in excellent agreement with the exact one and it reproduces
all of the structures at the correct energies and with the proper
magnitude. Aside from a small overshoot of the curve between about
0.3 and 0.4 eV due to the derivative technique which is known to
produce such characteristics just after the end of the boson
spectrum, the op-qp curve also crosses the bare dispersion at the
same point as the qp curve. The agreement is remarkable and
similar agreement is found in the detailed comparison of the real
and imaginary parts of $\Sigma^{qp}$ and $\Sigma^{op-qp}$ (not
shown). The grey shading shows the broadening of the dispersion
curve which occurs with $\Sigma_2$ and is indicated by
$\pm|\Sigma_2^{qp}|$. The inset of this figure shows the procedure
of Eq.~(\ref{eq:sigopqp}) applied to experimental optical data for
optimally-doped Bi-2212 at $T=100K$ \cite{timusk,bi2212opt}.
Again, $\Sigma^{op}$ gives a featureless dispersion curve, but
$\Sigma^{op-qp}$ restores structures in the curve (such as the
kink at $\sim 50$ meV) which can now be compared to ARPES.
Fig.~\ref{Fig1} emphasizes that it is $\Sigma^{op-qp}$ which must
be used for this purpose. A point that must be made before
continuing to such a comparison is that optics is a transport
property and it does not contain momentum-dependent information.
Consequently, along with the assumption that
$\alpha_{tr}^2F=\alpha^2F$, we stress that the ``dispersion
curves'' presented here are to be viewed as momentum-averaged
curves in terms of the self-energy and the only remaining
momentum-dependence enters through the bare dispersion
$\epsilon_k$ which is taken from LDA calculations.
%
%
\begin{figure}[t]
  \vspace*{-0.0 cm}%
  \centerline{\includegraphics[width=3.00 in]{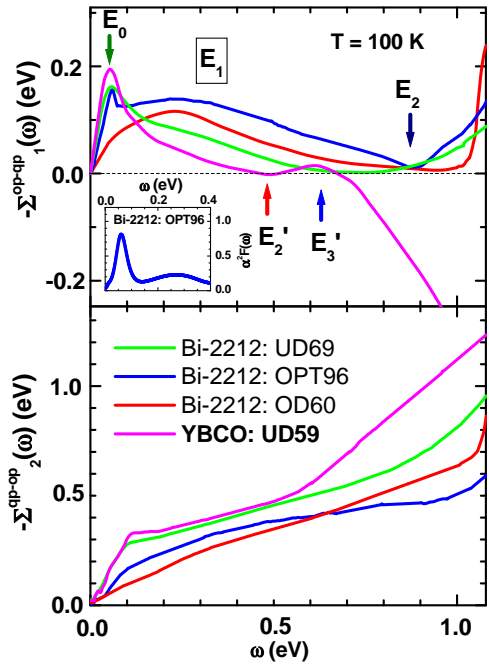}}%
  \vspace*{-1 cm}%
\caption{(color online) Upper frame: The real part of the
optically-derived quasiparticle self-energy for three Bi-2212
samples, overdoped, optimally-doped and underdoped, and one
underdoped YBCO in the orthoII phase~\cite{orthoII}, all at
$T=100$ K. Lower frame: The imaginary part of the
optically-derived quasiparticle self-energy. Various energy scales
are indicated on the curves and discussed in the text. The inset
gives the $\alpha^2F(\omega)$ obtained in the optimally doped
Bi-2212 from inversion of optical data.~\cite{hwang06}}
 \label{Fig3}
\end{figure}

Having established the technique, we now present our new results
on high energy scales in Bi-2212 and YBCO as imaged by optics. In
Fig.~\ref{Fig2}, we present the analysis of the optical data for
optimally-doped Bi-2212 at 100 K. The derivative needed in
Eq.~(\ref{eq:sigopqp}) to extract $\Sigma^{op-qp}$ can introduce
noise and we have applied smoothing techniques as is reasonable
since we are interested mainly in capturing global features. For
comparison with ARPES, we have presented the $\Sigma^{op-qp}$ in
the form of a dispersion curve using an LDA
calculation\cite{bansil} for the bare dispersion $\epsilon_k$.
With this information, we can form a color map of the
quasiparticle spectral function $A(k,\omega)=\frac{1}{\pi}
|\Sigma_2^{op-qp}|/
[(\omega-\epsilon_k-\Sigma_1^{op-qp})^2+(\Sigma_2^{op-qp})^2]$ to
connect to the raw data as presented by ARPES. In
Fig.~\ref{Fig2}b, we compare our data with that of Graf et
al.\cite{graf} from ARPES in the nodal direction on a similar
optimally doped sample. In the ARPES, they identify three energy
scales with these data at $E_0\sim 50$ meV, $E_1\sim 0.4$ eV, and
$E_2\sim 0.8$ eV. In our data, we can confirm a structure at
$E_2\sim 0.9$ eV and a much stronger kink at $E_0\sim 50$ meV,
however, we do not find evidence for a very strong feature at
$\sim 0.4$ eV, although the $\Sigma_1^{op-qp}$ does show a local
maximum around 0.25 eV. The broadening that we find is within the
range that is found by ARPES. Thus, we conclude that overall, we
find agreement with ARPES using our analysis as shown in
Fig.~\ref{Fig2}b, however, there remain some discrepancies which
we will now address.

Inversion\cite{schachinger,carbnat} of optical data to recover the
electron-boson spectral density of Eqs.~(\ref{eq:tauapprox}) and
(\ref{eq:lamapprox}) shows that the kink at $\sim 50$ meV may be
attributed to a peak in $\alpha^2 F(\omega)$ which is followed by
a broad spin fluctuation background extending beyond 0.4 eV (see
inset of Fig.~\ref{Fig3}). This spectrum accounts for the
deviations, extending up to $\sim$ 0.6 eV, of the op-qp data from
the LDA curve seen in Fig.~\ref{Fig2}b. The renormalization found
in the ARPES data, while larger than that found in the op-qp curve
around 0.4 eV, is smaller below the 50 meV kink. The lack of
quantitative agreement between the optics and ARPES in these two
regions may reflect (i) the differences between the quasiparticle
and transport spectral densities, not included in this work, (ii)
the differences in k-dependence between optical data (averaged
over the Fermi surface) and that of ARPES (momentum specific), and
(iii) that formally the self-energy $\Sigma(\bf{k},\omega)$ can
depend on the magnitude of $\bf{k}$ as well as direction, but this
goes beyond the theory used here. It is clear that
renormalizations are present in both experiments on a broad energy
scale up to the end of the spin fluctuation spectrum ($\sim$ 0.4
eV) which we take here to be the $E_1$ scale rather than the
position of the peak in the self-energy. Furthermore, we associate
the higher energy scale around 0.8-0.9 eV with the commencement of
a new higher energy band (see discussion below). One of the
results, strongly highlighted in some of the ARPES
work\cite{vallas06,graf06,pan06} is the vertical dispersion seen
in the data from $E_1$ to $E_2$. The question arises as to whether
this is due to some sort of ordering\cite{graf,graf06} or if it is
a renormalization effect. Indeed a vertical dispersion can be
obtained within standard electron-boson
models\cite{dogan,cappelluti} with appropriate choice of
parameters and in this instance, the high energy scale of this
feature points to a spin-fluctuation spectrum rather than a low
energy boson spectrum. Thus, we would conclude that the large kink
is due to renormalization rather than something more exotic, as
was also concluded by Valla et al.\cite{vallas06}.

In the upper frame  of Fig.~\ref{Fig3}, we show $\Sigma_1^{op-qp}$
for three Bi-2212 materials: underdoped (UD69), optimally-doped
(OPT96) and overdoped (OD60), and one underdoped YBCO material
(UD59)~\cite{orthoII}, all at $T=100$ K in the normal state. For
the Bi-2212 samples we have indicated possible features that would
correlate with the $E_0$, $E_1$, and $E_2$ seen in ARPES providing
possible confirmation of three similar energy scales seen
universally across members of Bi-2212 family. For the YBCO sample,
we also can suggest the existence of the three energy scales.
While $E_0$ remains the same, the $E_1$ and $E_2$ scales are now
shifted to lower energy relative to Bi-2212 and are labelled
$E_1'$(not shown) and $E_2'$. In addition, there is a possible
fourth scale $E_3'\sim 0.6$ eV. In the lower frame, we show
$-\Sigma_2^{op-qp}$ for the four samples. This quantity
 is positive definite and shows some of the
characteristic energy scales more unambiguously. In particular, we
see the emergence of a new band at larger energy. In the case of a
finite band, we expect the scattering rate to
drop~\cite{antonclean,antonimp,cappelluti} beyond the end of the
boson spectrum.

In summary, we have demonstrated a new analysis to extract the
quasiparticle self-energy directly from the optical conductivity
data, which will facilitate future comparisons with both ARPES and
theory. In applying this method to Bi-2212 for various dopings, we
find good agreement with the ARPES data in comparison with the
optically-derived quasiparticle dispersion curve. While the optics
data show features at both low and high energies ($E_0\sim 50$ meV
and $E_2\sim 0.9$ eV) as in ARPES, the profile of the intermediate
scale $E_1\sim 0.4$ eV, which has been identified as the
background spin fluctuation scale, is very different in optics.
Notwithstanding the lack of momentum resolution, optics provides a
complementary technique to ARPES in this case because it is a bulk
probe, has enhanced energy resolution, and is able to examine a
wider range of materials such as the YBCO family. Indeed, we find
with the YBCO data shown here that we can identify up to four
energy scales, which range from low to high energies, but that
overall they appear to be shifted to lower energy relative to
Bi-2212. The theoretical understanding of these features remains
to be more fully elucidated, but confirmation of these scales by
more than one technique gives impetus to such an endeavour.

This work has been supported by the Natural Science and
Engineering Council of Canada (NSERC) and the Canadian Institute
for Advanced Research (CIAR).


\begin{thebibliography}{99}

\bibitem{campuzano} C. Campuzano {\it et al},
in {\it Physics of Conventional and Unconventional
Superconductors}, Edited by K. H. Bennemann and J. B. Ketterson
(Springer-Verlag, Berlin, 2003) vol. 2, p. 167.

\bibitem{damascelli} A. Damascelli {\it et al.},
Rev. Mod. Phys. {\bf 75}, 473 (2003).

\bibitem{basovrmp} D. N. Basov {\it et al.},
{\bf 77}, 721 (2005).

\bibitem{zhou} X.J. Zhou {\it et al.},
Phys. Rev. Lett. {\bf 95}, 117001 (2005).

\bibitem{verga} S. Verga {\it et al.},
Phys. Rev. B {\bf 67}, 054503 (2003).

\bibitem{marsiglio} F. Marsiglio {\it et al},
Phys. Lett. A {\bf 245}, 172 (1998).

\bibitem{dordevic} S.V. Dordevic {\it et al.},
Phys. Rev. B {\bf 71}, 104529 (2005).

\bibitem{schachinger} E. Schachinger {\it et al.},
Phys. Rev. B {\bf 73}, 184507 (2006).

\bibitem{carbnat} J.P. Carbotte {\it et al.},
Nature (London) {\bf 401}, 354 (1999).

\bibitem{timusk} J. Hwang {\it et al.},
Nature (London) {\bf 427}, 714 (2004).

\bibitem{hwang06} J. Hwang {\it et al.},
cond-mat/0610127 (2006).

\bibitem{kaminski} A. Kaminski {\it et al.},
Phys. Rev. Lett. {\bf 86}, 1070 (2001).

\bibitem{kaminski2} A. Kaminski {\it et al.},
Phys. Rev. Lett. {\bf 84}, 1788 (2000).

\bibitem{sato} T. Sato {\it et al.},
Phys. Rev. Lett. {\bf 91}, 157003 (2003).

\bibitem{kordyukmag} A.A. Kordyuk {\it et al.},
Phys. Rev. Lett. {\bf 92}, 257006 (2004).

\bibitem{kordyuk2} A.A. Kordyuk {\it et al.},
Phys. Rev. Lett. {\bf 97}, 017002 (2006).

\bibitem{kim} T.K. Kim {\it et al.},
Phys. Rev. Lett. {\bf 91}, 167002 (2003).

\bibitem{johnson} P.D. Johnson {\it et al.},
Phys. Rev. Lett. {\bf 87}, 177007 (2001).

\bibitem{lanzara} A. Lanzara {\it et al.},
Nature (London) {\bf 412}, 510 (2001).

\bibitem{cuk} T. Cuk {\it et al.},
Phys. Rev. Lett. {\bf 93}, 117003 (2004).

\bibitem{graf} J. Graf {\it et al.},
cond-mat/06077319 (2006).

\bibitem{xie} B.P. Xie {\it et al.},
cond-mat/0607450 (2006).

\bibitem{vallas06} T. Valla {\it et al.},
cond-mat/0610249 (2006).

\bibitem{graf06} J. Graf {\it et al.},
cond-mat/0610313 (2006).

\bibitem{pan06} Z.-H. Pan {\it et al.},
cond-mat/0610442 (2006).

\bibitem{antonclean} A. Knigavko {\it et al.},
Phys. Rev. B {\bf 72}, 035125 (2005).

\bibitem{antonimp} A. Knigavko {\it et al.},
Phys. Rev. B {\bf 73}, 125114 (2006) .

\bibitem{cappelluti} E. Cappelluti {\it et al.},
Phys. Rev. B {\bf 68}, 224511 (2003).

\bibitem{boza} B. Mitrovi\'c {\it et al.}
Phys. Rev. B {\bf 20}, 6749 (1984).

\bibitem{allen} P.B. Allen, Phys. Rev. B {\bf 3} 305 (1971).

\bibitem{shulga} S.V. Shulga {\it et al},
Physica C {\bf 178}, 266 (1991) .

\bibitem{sharapov} S.G. Sharapov {\it et al},
Phys. Rev. B {\bf 72}, 134506 (2005) .

\bibitem{carbotte} J.P. Carbotte {\it et al.},
Phys. Rev. B {\bf 71}, 054506 (2005).

\bibitem{schachinger03} E. Schachinger {\it et al.},
Phys. Rev. B {\bf 67}, 214508 (2003).

\bibitem{bi2212opt} J. Hwang {\it et al.},
cond-mat/0607653 (2006).

\bibitem{bansil} Hsin Lin {\it et al.},
Phys. Rev. Lett. {\bf 96}, 097001 (2006).

\bibitem{dogan} F. Dogan {\it et al.},
cond-mat/0603635 (2006).

\bibitem{orthoII} J. Hwang {\it et al.},
Phys. Rev. B {\bf 73}, 014508 (2006).

\end{thebibliography}
\end{document}